\documentclass[11pt,twoside,a4paper]{article}
\usepackage{epsfig}
\usepackage{graphicx}
\usepackage{srt_style}
\begin{document}
\baselineskip .5cm
%
%%%%  Title of your paper %%%%%%
\title{Single-dish monitoring of circumstellar water masers}

%%%%  The author(s), separated by commas; do not put a comma before the last author  %%%%
\author{J. Brand$^1$, L. Baldacci$^{1,2}$, D. Engels$^3$}

%%%%  Affiliation of the author(s)   %%%%%%
% Put here your coordinates, including the complete mailing address of your home Institute
% and your email address
\affil{1. Istituto di Radioastronomia, C.N.R., Via Gobetti 101, I-40133 
Bologna, Italy (brand@ira.cnr.it)\\
2. Osservatorio Astronomico, Via Ranzani 1, I-40127 
Bologna, Italy (baldac\_s@ira.cnr.it)\\
3. Hamburger Sternwarte, Gojensbergweg 112, D-21029 Hamburg, 
Germany (dengels@hs.uni-hamburg.de)}

% Write the text starting from here and using the usual LaTeX commands.
%
\begin{abstract}
We present an overview of the long-term H$_2$O maser monitoring program of a
sample of late-type stars, carried out with the Medicina 32-m and Effelsberg 
100-m telescopes, and
describe the results in some detail. The role the SRT could play in this 
program is outlined. 
\end{abstract}

\section{Introduction}
Maser emission from the 6$_{16} - 5_{23}$ rotational transition of water at
22.2~GHz is a common feature in both circumstellar shells and in star-forming
regions. In both types of sources the maser emission is highly variable. For
the masers associated with Young Stellar Objects this variation is mostly 
erratic, while those associated with late-type stars sometimes vary in phase 
with the luminosity of the central star, and at other times may show highly 
irregular behaviour, including spectacular flaring events.

\section{Monitoring water masers in circumstellar envelopes} 
Over the past decade we observed a sample of about 20 late-type stars 
(supergiants, semi-regular (SR) variables, OH/IR stars, and Miras) $3-4$ times 
per year in the 1.3-cm line of H$_2$O. The observations were carried out with 
the Medicina 32-m and Effelsberg 100-m telescopes. In addition, a sub-sample of
these stars was observed with the ISO satellite, and at several epochs with 
the VLA. In our analysis we include data from the literature.
The aim is to investigate the maser variability as a function of
both time and stellar parameters (such as optical/IR variability, mass loss 
rate, spectral type, IR colours).

\smallskip\noindent
The extensive single-dish observing program was started in 1990, and continues 
up to the present day. Including observations in the Arcetri archives, that go
back to 1987, for several stars we have (Medicina) data over a 15-year
time span. To our knowledge this constitutes the largest continuous monitoring 
data base, second only to that of the Pushchino Observatory; such a long 
time-coverage is a necessity especially for the study of OH/IR stars, which 
can have periods of up to 3000~days.
The table lists the stars which have been most frequently monitored 
in our program.

\begin{table}
\caption[]{Most frequently monitored stars}
\begin{tabular}{l|l}
\hline
Stellar type& Name    \\
\hline
Mira        & o\,Cet, IK\,Tau, R\,Leo, U\,Her, RR\,Aql, R\,Cas \\
Semi-Regular& RT\,Vir, RX\,Boo, SV\,Peg \\
OH/IR       & IRC+10011, OH26.5+0.6, OH32.8$-$0.3, OH39.7+1.5, \\
            & OH44.8$-$2.3, OH83.4$-$0.9 \\ 
Supergiant  & VX\,Sgr, NML\,Cyg \\
\hline
\end{tabular}
\end{table}

\section{Methods of visualization \& analysis}
One of the biggest problems in dealing with a data base of this size, where 
variations in time {\it and} velocity {\it and} intensity are of importance, 
is to find a compact way to represent the data, while allowing variations in all
three parameters to be visible at a glance. An elegant way to do this 
is shown in Fig.~1. 
As a first approach in the analysis the data are presented in various 
graphical forms which allow one to follow the variation of the different 
spectral components. In particular, five quantities have been considered: 
the total velocity range of the maser emission; the intensity of the maser 
lines as a function of time and velocity; the flux density integrated over 
the observed velocity range as a function of time; the maximum (minimum) flux 
density ever observed at each velocity; and the frequency of 
occurrence of the maser emission at any given velocity. 

\smallskip\noindent
Fig.~2a shows the so-called {\it upper envelope} of the emission, 
which is the maximum flux density ever observed in each channel and shows
what the maser spectrum would look like if all components emitted at their 
maximum output simultaneously, and provides a measure for the maximum maser
luminosity $\rm L_{H_2O}(up)$, as well as the intensity-weighted mean velocity 
V$_{up}$ and velocity dispersion $\Delta$V$_{up}$. Likewise, 
the {\it lower envelope} (not shown) identifies the components that 
were always present during the monitoring. Fig.~2b is an example of a 
{\it frequency histogram}, showing the number of times that emission has been 
detected at
each velocity in the spectrum. From the histogram a mean velocity V$_{fr}$ and
dispersion $\Delta$V$_{fr}$ are derived. Both V$_{up}$ and V$_{fr}$ are 
useful to separate the red- and blue-shifted velocity features in a consistent
way. While V$_{up}$ is more susceptible to the presence of very strong peaks,
V$_{fr}$ is more influenced by groups of emission lines that are present for
long periods of time (and therefore expected to be less useful for objects 
with stable emission at velocities well away from the stellar velocity).

\noindent
An important parameter is the total flux ${\rm S_{tot} = \int F(H_2O) d\nu}$ 
and its behaviour with time. Especially for the 
Miras in the sample this can be used to derive the period. 
For the interesting case of R\,Cas, see Brand et al.~\cite{jbbrasil}. The 
total velocity range of the maser is another quantity which may vary with the 
optical period of the star. Where possible, we identified individual emission
components in the spectra, and followed their behaviour (in flux density, 
velocity, and linewidth) with time. The variation of flux density with time
mostly reflects the stellar variability, and usually the components reach the 
same maximum value each period, but sometimes strong flares are found. 
In many cases the variation of velocity indicates acceleration or deceleration 
of the individual maser components, although blending of components potentially
complicates this interpretation.

\begin{figure}
\centerline{\includegraphics[width=12cm]{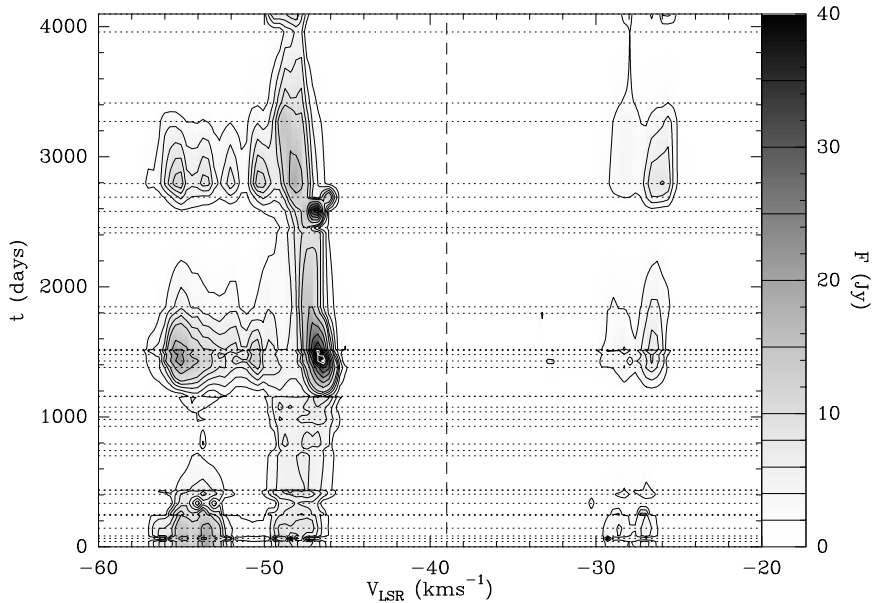}}
\caption{
Grey-scale plot and contour map of the H$_2$O flux density
versus velocity as a function of time for OH83.4$-$0.9. The horizontal dotted
lines indicate when a spectrum is available; observations separated by less
than 10~days were averaged. Between adjacent observations linear interpolation
was performed. Contour values (start(step)end) are 2(2)10(5)40~Jy (black);
45(5)55~Jy (white). The vertical dashed line indicates the stellar velocity;
t=0 at JD2447940.}
\end{figure}

\begin{figure}
\centerline{{\rotatebox{-90}{\includegraphics[height=12cm]{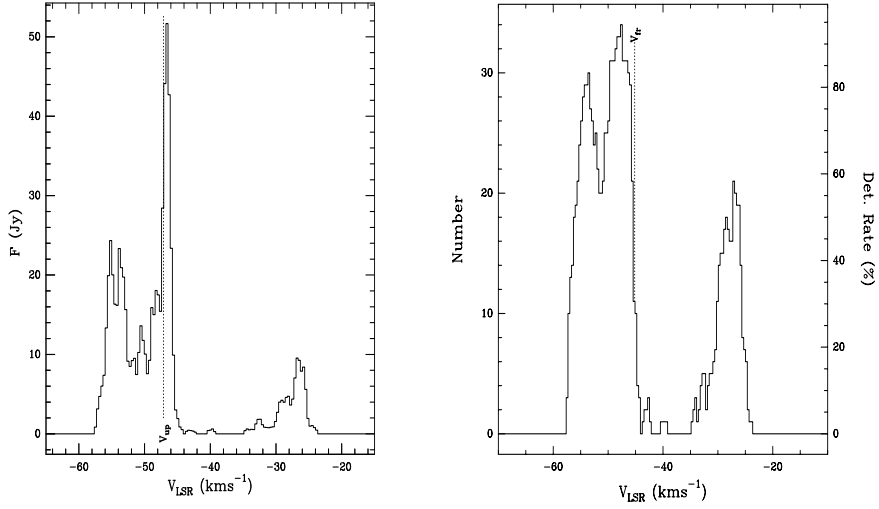}}}}
\caption{
{\bf a.}\ (left) Upper envelope of the H$_2$O emission 
(above $5\,\sigma$) of OH83.4$-$0.9.
{\bf b.}\ (right) Frequency histogram of OH83.4$-$0.9. Shown for each channel 
is the number of times that the emission reached above the 5\,$\sigma$-level 
during the entire monitoring period.\hfill\break\noindent
For both panels, the data were resampled to a resolution of 0.33~kms$^{-1}$; 
V$_{up}$ and V$_{fr}$ (see text) are indicated.
}
\end{figure}

\section{Results}
Because of space restrictions, we only list some of the main results 
of the analysis of 10 stars from our sample.

\smallskip\noindent
$\bullet$\ The brighter the central star (larger L$_{bol}$), the stronger the
maser (larger L$_{H_2O}$) (Fig.~3a).

\smallskip\noindent
$\bullet$\ Stronger masers have more components, and show less variability 
(Figs.~3b, c). 
Stronger masers are also characterized by larger mass loss and larger shell 
size.

\smallskip\noindent
$\bullet$\ Mass loss rate \.{M} and the size of the circumstellar envelope are
correlated. In Miras and SR-variables (smaller \.{M}) the
water masers originate from regions relatively close to the star, and have
preferentially tangential gain paths; the velocity range of the emission is
$\sim 15$~kms$^{-1}$. It is roughly twice that for the OH/IR and Supergiant
stars (higher \.{M}), where the masers are found at larger distances from the
star and radial gain paths dominate. The transition between the two regimes
seems to occur at \.{M} $\sim 4 \times 10^{-6}$~M$_{\odot}$yr$^{-1}$ (Fig.~3d).

\smallskip\noindent
$\bullet$\ Many emission components in the spectra show a change in V$_{lsr}$
with time. Where blending does not seem to be a problem, values of
(de-)acceleration between 0.06 and 0.40~kms$^{-1}$yr$^{-1}$ are typical; the
largest velocity change is found for a flare component in RX\,Boo:
$-$0.4~kms$^{-1}$ in 84~days (=$-1.74 \pm 0.20$~kms$^{-1}$yr$^{-1}$).

\smallskip\noindent
$\bullet$\ For stars where a period could be determined from the maser data, we
find that the period of the maser is the same as that of the optical and IR
emission. There is however a phase delay ($\sim 0.1-0.3$) for the H$_2$O maser.

\smallskip\noindent
$\bullet$\ The emission, integrated over the blue (V$<V_{\ast}$) and red
(V$>V_{\ast}$) parts of the spectra, shows the same change with time.
The masers in OH/IR stars have radial gain paths, and in these objects the 
blue part of the emission dominates over the red part. This can be caused by
maser amplification of the stellar radio-continuum radiation and/or geometric 
blocking of the red light by the star (Takaba et al. \cite{takaba}).

\begin{figure}
\centerline{\includegraphics[width=14cm]{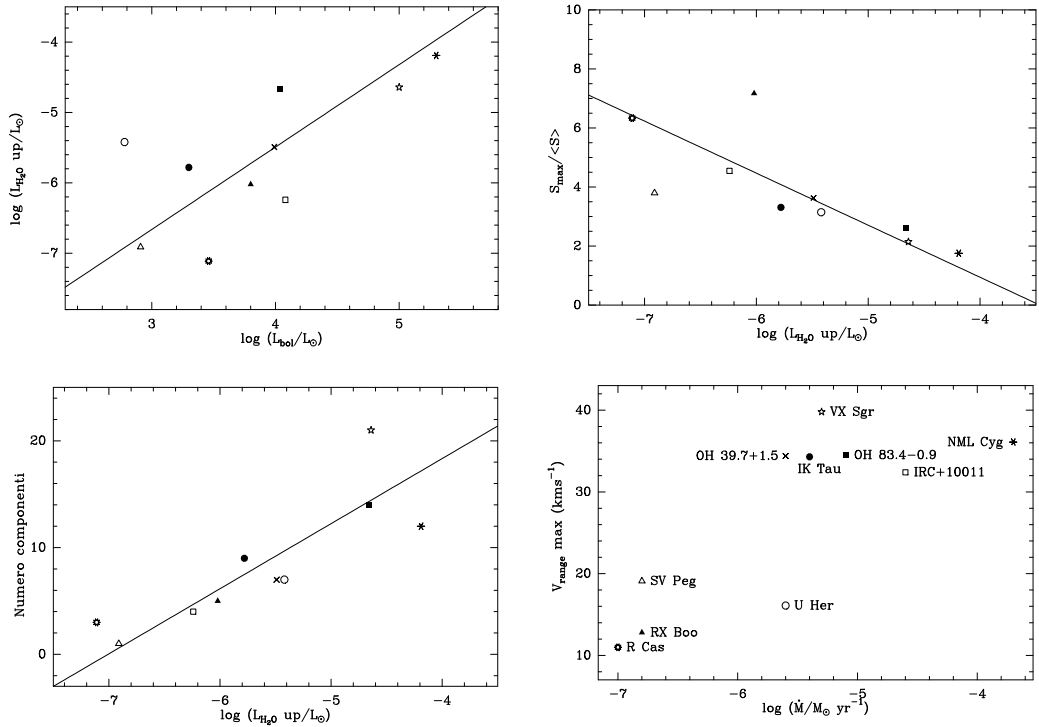}}
\caption{
{\bf a.}\ (top left) $\rm L_{H_2O}(up)$, the maximum maser luminosity,
determined from the {\it upper envelope}, as a function of the stellar
L$_{bol}$
{\bf b.}\ (top right) The variability index, defined as the ratio between
the maximum integrated flux density S$_{max}$ reached during the observations,
and the average flux density $<$S$>$, as a function of $\rm L_{H_2O}(up)$
{\bf c.}\ (bottom left) The number of emission components in the spectrum with
the highest $\rm \int Fd\nu$, as a function of $\rm L_{H_2O}(up)$
{\bf d.}\ (bottom right) The maximum detected velocity range of the maser
emission as a function of the mass loss.}
\end{figure}

\section{What can the SRT do for us?}
The arrival of the SRT will both improve the quality of the maser data, and
spawn new research projects. 
A direct consequence of the presence of the SRT will be that we can perform
more frequent monitoring of our objects, resulting in a better time-coverage.
The higher angular resolution of the SRT will reduce spectral contamination by 
nearby, unrelated water masers, which can be a problem for masers in 
star-forming regions. The expected superior quality of the surface of the 
dish results in higher sensitivity, a cleaner beam, and hence more efficient
observations, and the possibility to detect fainter maser components. We 
should also be able to observe more (faint) calibrators than
what is presently possible at Medicina. 

\smallskip\noindent
During the years of monitoring several flares were 
found in the maser spectra. The causes for these phenomena are not known, nor 
is their frequency of occurrence.
A systematic patrol, facilitated by the presence of the SRT, of a large sample 
of bright maser sources, followed by target-of-opportunity observations at 
the VLA (or VLBA), may result in a better understanding of maser flares.

\smallskip\noindent
Models of circumstellar shells have become increasingly complex, and involve
non-spherical multiple shells, which are the result of mass loss rates that
vary on time-scales of years or even decades. If these variations in mass loss
rates are common in late-type stars, one may be able to see the effects also
in the water maser properties (e.g. changes in average luminosity, and 
velocity range of the emission over the years). Continued and frequent 
monitoring of a large sample of objects is required to reveal these gradual
changes, and will help to constrain wind models.

\smallskip\noindent
Finally, the detectors planned for the SRT will allow observations of other 
masers as well, such as those of SiO.

\end{document}